\documentclass[conference,a4paper]{APSIPA2026}
\usepackage{amsmath}
\usepackage{graphicx}
\usepackage{booktabs}
\usepackage{multirow}
\usepackage{threeparttable}
\usepackage{subcaption}
\usepackage{balance}
\usepackage[backend=biber,style=ieee,]{biblatex}
\usepackage{geometry}
\usepackage{fancyhdr}

\fancypagestyle{firststyle}{
  \fancyhf{}
  \fancyhead[C]{2026 Asia Pacific Signal and Information Processing Association Annual Summit and Conference (APSIPA ASC)}
}

\newcommand{\Lkd}{\ensuremath{\mathcal{L}_{\text{KD}}}}
\newcommand{\Lcl}{\ensuremath{\mathcal{L}_{\text{CL}}}}

\begin{document}

\title{Correlation-Based Distillation Yields More Mergeable Speech-Music Encoders}

\author{
\authorblockN{
Fabian Ritter-Gutierrez\authorrefmark{1}
}

\authorblockA{
\authorrefmark{1}
Nanyang Technological University, Singapore \\
E-mail: s220064@e.ntu.edu.sg}
}

\maketitle
\thispagestyle{firststyle}
\pagestyle{empty}

\begin{abstract}
A single compact encoder for both speech and music removes the need to maintain a separate model per domain. A practical recipe distils a speech teacher and a music teacher into two small students, then merges them into one model. Two merging algorithms exist: interpolating the student weights from a shared initialisation, or permuting their channels into alignment before averaging. Prior work fixed the distillation loss in both, leaving open whether the loss itself affects how well the students merge. We show that it does. Holding architecture and evaluation fixed, we replace the DistilHuBERT loss \Lkd{} with a correlation-based loss \Lcl{}. Under weight interpolation, \Lcl{} students stay closer to their own better endpoint in $17$ of $18$ comparisons and lead on the speech tasks at every interpolated weight. Under activation permutation, they need fewer channels rearranged at three layers, their matched channels correlate more strongly, and they again lead on speech. The two merging algorithms share no mechanism, yet both improve under the same loss substitution. This suggests that the gain lies in the representation \Lcl{} produces rather than in either algorithm.
\end{abstract}


\section{Introduction}
\label{sec:intro}

A single small encoder that represents both speech and music is desirable for applications that need general audio representations, e.g.\ audio large language models~\cite{salmon}. Pre-training from scratch on both domains is computationally expensive. A more practical route is knowledge distillation~\cite{hintonDistill} followed by post-hoc merging, in which a speech self-supervised learning (SSL) teacher and a music SSL teacher are distilled into compact students that are then combined into one model without retraining.

Two post-hoc merging tools were introduced in prior work. Task arithmetic (TA)~\cite{task_arithmetic,ritter_ta_is2025} interpolates task vectors
of distilled students from a shared initialisation but never varied the distillation loss.
Correlation-permutation (CP)~\cite{correlation_permutation_ritter} is a training-free activation-alignment method applied only to
the large pre-trained encoders, leaving open whether it extends to the distillation setting. Neither study addressed whether the choice of distillation
loss changes the mergeability of the resulting students. If so, the merge-time behaviour reported for any single merging algorithm is partly a property of the student,
not of the algorithm.

The correlation-based distillation objective \Lcl{} was introduced for noise-robust SSL distillation~\cite{ritter_noise_robust}.
Here we treat \Lcl{} as a given training objective and study its merge-time behaviour. We substitute \Lcl{} for the standard distillation
loss \Lkd{}, hold the students and a fixed nine-task evaluation constant, and apply two merging algorithms that share no mechanism.
One operates in weight space from a shared initialisation; the other aligns activations of their representations.
If the loss were merge-neutral, substituting \Lcl{} for \Lkd{} would leave both algorithms unchanged.

In this paper we show the following. (1)~Under task arithmetic, \Lcl{} stays closer to its own better endpoint in $17$ of the $18$ domain-by-weight-by-size comparisons, and gives the higher Speech Score at every interpolated weight. (2)~We extend correlation-permutation from pre-trained encoders to distilled students. \Lcl{} students need less channel rearrangement at three layers, admit stronger channel matches, and lead on Speech Score at both weights and depths. (3)~Two merging algorithms that share no mechanism order the two losses the same way, which points to the representation \Lcl{} produces rather than to either algorithm.

Task arithmetic and the \Lkd{} students come from~\cite{ritter_ta_is2025}, correlation-permutation from~\cite{correlation_permutation_ritter}, and \Lcl{} from~\cite{ritter_noise_robust}. That task-arithmetic study used \Lkd{} only. New here are all \Lcl{} merges under both algorithms, all permutation results and alignment statistics on distilled students, the wide configuration and the nine-task re-scoring.

\section{Background}
\label{sec:background}

This section recaps the two merging tools and the distillation loss used in the protocol of Section~\ref{sec:protocol}. All three are prior work.

\textbf{Task arithmetic (TA).} A speech teacher (HuBERT~\cite{hubert}) and a music teacher (MERT~\cite{mert}) are each distilled from a shared initialisation $\boldsymbol{\theta}_0$, giving student weights $\boldsymbol{\theta}_S$ and $\boldsymbol{\theta}_M$. The speech and music vectors are $\boldsymbol{\tau}_S = \boldsymbol{\theta}_S - \boldsymbol{\theta}_0$ and $\boldsymbol{\tau}_M = \boldsymbol{\theta}_M - \boldsymbol{\theta}_0$. The merged encoder interpolates them as~\cite{ritter_ta_is2025,task_arithmetic,model_soup}
\begin{equation}
\label{eq:ta}
\boldsymbol{\theta}_{\text{merged}} = \boldsymbol{\theta}_0 + \lambda_1\, \boldsymbol{\tau}_S + \lambda_2\, \boldsymbol{\tau}_M,
\qquad \lambda_2 = 1 - \lambda_1,
\end{equation}
where $\lambda_1 \in [0,1]$ trades speech against music at merge time.

\textbf{Correlation-permutation (CP).} Two independently trained encoders need not share a coordinate system, so a permutation can be applied before averaging~\cite{linear_inter_lottery_hyp,permutation_linear_interpolation,git-rebasin}. CP~\cite{correlation_permutation_ritter} solves a per-layer linear-assignment problem maximising channel-wise cross-correlation between the two encoders' activations. The permutation covers attention heads, within-head channels, the second feed-forward network (FFN) projection and the convolutional neural network (CNN) front end, leaving the query, key and value projections untouched. CP is training-free, needs about ten thousand calibration samples, and requires no shared initialisation. With $\boldsymbol{\pi}^{*}$ the per-layer permutations, the merged encoder is
\begin{equation}
\label{eq:cp}
\boldsymbol{\theta}_{\text{merged}} = \lambda_1\, \boldsymbol{\theta}_S + (1-\lambda_1)\, \boldsymbol{\pi}^{*}(\boldsymbol{\theta}_M),
\end{equation}
so only the music student moves. An inverse permutation at the next layer's input makes $\boldsymbol{\pi}^{*}(\boldsymbol{\theta}_M)$ function-preserving, so $\lambda_1 \in \{1, 0\}$ recovers the two students and $\lambda_1$ means the same as in~(\ref{eq:ta}).

\textbf{The distillation loss \Lcl{}.} The standard distillation objective \Lkd{} is the DistilHuBERT layer-prediction loss~\cite{distilhubert}. The correlation-based objective~\cite{ritter_noise_robust} replaces it with
\begin{equation}
\label{eq:lcl}
\Lcl{} = \mathcal{L}_{\text{CC}} + \mathcal{L}_{\text{SC}} - \gamma\, \mathcal{L}_{\cos}.
\end{equation}
The cross-correlation term drives the Pearson correlation matrix $\mathbf{C}_{\text{cc}}$ between student and teacher features towards identity:
\begin{equation}
\label{eq:lcc}
\mathcal{L}_{\text{CC}} \!=\! \sum_{p,t}\!\Bigg[\sum_{i}\!\big(1{-}(\mathbf{C}_{\text{cc}})_{ii}\big)^{\!2} + \lambda_{\text{cc}}\!\!\sum_{i \neq j}\!(\mathbf{C}_{\text{cc}})_{ij}^{2}\Bigg],
\end{equation}
rewarding diagonal alignment and penalising cross-dimensional coupling. The self-correlation term decorrelates the student's features via the off-diagonal of $\mathbf{C}_{\text{sc}}$:
\begin{equation}
\label{eq:lsc}
\mathcal{L}_{\text{SC}} = \lambda_{\text{sc}} \sum_{p,t} \sum_{i \neq j} (\mathbf{C}_{\text{sc}})_{ij}^{2}.
\end{equation}
$\mathcal{L}_{\cos}$ retains the DistilHuBERT cosine term, scaled by $\gamma$. The objective belongs to the redundancy-reduction family~\cite{barlowtwins}; its noise-robustness mechanism is analysed in~\cite{ritter_noise_robust}.

\textbf{Ensemble distillation.} Rather than merging post-hoc, one may distil both teachers simultaneously into a single student~\cite{multi_distill_asru2025,distilhubert}. The student receives gradients from both teachers during training, so no post-hoc interpolation is needed. We include this method as a baseline; its score is $\lambda$-invariant.

\section{Mergeability Protocol}
\label{sec:protocol}

The question in Section~\ref{sec:intro} is answerable only if the loss is the sole variable.
We therefore apply the same pool of shared-initialisation students to both merging algorithms and score them under one nine-task protocol, so the \Lkd{}-versus-\Lcl{} swap is the only difference.

The protocol follows~\cite{ritter_ta_is2025}. The four speech tasks come from SUPERB~\cite{superb,librispeech}: automatic speech recognition (ASR), keyword spotting (KS), intent classification (IC), and emotion recognition (ER). The four music tasks come from MARBLE~\cite{marble}: singer, vocal-technique, instrument, and genre identification. The ninth is environmental sound classification on ESC-50~\cite{esc50}. Three student configurations are tested: a two-layer transformer ($D{=}768$, ${\sim}23$\,M parameters), a three-layer transformer (${\sim}31$\,M), and a wide two-layer transformer ($D{=}1536$, feed-forward dimension $6144$, $6$ attention heads, ${\sim}62$\,M). The wide students start from a shared random checkpoint instead of HuBERT, removing the initialisation confound. The ensemble-distillation model of Section~\ref{sec:background} serves as the baseline comparison at each size.

Each task score is normalised between an FBANK floor and the better teacher, following~\cite{ritter_ta_is2025,correlation_permutation_ritter,ml_superb}:
\begin{equation}
\label{eq:score}
\text{Score}_u = \frac{1000}{|T|} \sum_{t \in T} \frac{s_t(u) - s_t(\text{FBANK})}{s_t(\text{BEST}) - s_t(\text{FBANK})},
\end{equation}
where $u$ is the model under evaluation, $s_t(u)$ its score on task~$t$, and $s_t(\text{BEST})$ the better teacher on that task. The Speech Score averages the four speech tasks, the Music Score the five music and audio tasks, and the Average Score the two domain scores. Robustness is measured as the deviation from the better single-domain endpoint~\cite{ritter_ta_is2025},
\begin{equation}
\label{eq:delta}
\Delta_t(\lambda) = s_t(\lambda) - s_t^{\text{best}},
\quad
s_t^{\text{best}} = \max\!\big(s_t(\lambda{=}1.0),\, s_t(\lambda{=}0.0)\big),
\end{equation}
with the sign reversed for ASR so that a negative $\Delta$ always denotes degradation. Mean $\Delta$ is reported across speech tasks and across music tasks, in percentage points (pp).

We use \emph{mergeable} in three senses that need not move together. \emph{Interpolation robustness} is the endpoint-relative mean $\Delta$ of~(\ref{eq:delta}), which ignores pre-merge performance. \emph{Structural alignment} is how many channels the permutation moves and how well they then match. \emph{Merge quality} is the Average Score of~(\ref{eq:score}).

\begin{table}[t]
\centering
\caption{Configuration shared by every student. Within a pair the students differ only in the teacher and, where stated, the loss. Source: distillation configuration of~\cite{ritter_ta_is2025}; permutation settings of~\cite{correlation_permutation_ritter}.}
\label{tab:setup}
\footnotesize
\setlength{\tabcolsep}{4pt}
\begin{tabular}{@{}ll@{}}
\toprule
Component & Setting \\
\midrule
Distillation data & LibriSpeech 960\,h, both students \\
Teachers & HuBERT Base; MERT (\texttt{MERT-v0-public}) \\
Optimiser & Adam, $2{\times}10^{-4}$, 10k linear warm-up \\
Batch / steps & $24$ / $200$k \\
Checkpoint & lowest pre-training loss on dev-clean \\
\Lcl{} weights & $\lambda_{\text{cc}}{=}5{\times}10^{-5}$, $\lambda_{\text{sc}}{=}5{\times}10^{-6}$, $\gamma{=}1$ \\
\midrule
CP calibration & 10k clips: 5k Music4All, 5k LibriSpeech-100 \\
CP scope & CNN $+$ heads $+$ within-head $+$ $\mathbf{W}_2$ \\
CP merge points & $13$ (2-layer), $16$ (3-layer) \\
\bottomrule
\end{tabular}
\end{table}

Table~\ref{tab:setup} lists the settings, which fix architecture, initialisation, data and step count across the \Lkd{}-versus-\Lcl{} comparison.

\section{Analysis of \Lcl{} on Task-Arithmetic Interpolation}
\label{sec:ta}

Under task arithmetic the weight $\lambda_1$ controls the speech-music trade-off. Table~\ref{tab:ta_robust} reports the mean $\Delta$ of~(\ref{eq:delta}) as $\lambda_1$ moves inward. \Lcl{} is the more robust loss in $17$ of the $18$ comparisons across two domains, three weights and three sizes. The exception is speech at $\lambda_1{=}0.9$ at two layers ($-0.30$ against $-0.63$ pp). The margins are widest at three layers, at $\lambda_1{=}0.8$ on music ($-0.59$ against $-4.51$ pp) and on speech ($-1.01$ against $-3.01$ pp).

\begin{table}[t]
\centering
\caption{Mean deviation $\Delta$ of~(\ref{eq:delta}) from the better single-domain endpoint under task arithmetic (percentage points; closer to zero is more robust).}
\label{tab:ta_robust}
\footnotesize
\setlength{\tabcolsep}{4pt}
\begin{tabular}{@{}llcccccc@{}}
\toprule
 & & \multicolumn{2}{c}{2-layer} & \multicolumn{2}{c}{3-layer} & \multicolumn{2}{c}{Wide} \\
\cmidrule(lr){3-4}\cmidrule(lr){5-6}\cmidrule(lr){7-8}
Domain & $\lambda_1$ & \Lkd{} & \Lcl{} & \Lkd{} & \Lcl{} & \Lkd{} & \Lcl{} \\
\midrule
\multirow{3}{*}{Speech} & 0.9 & $\mathbf{-0.30}$ & $-0.63$ & $-0.78$ & $-0.33$ & $-0.41$ & $-0.40$ \\
                        & 0.8 & $-2.33$ & $-1.80$ & $-3.01$ & $\mathbf{-1.01}$ & $-1.56$ & $-1.38$ \\
                        & 0.5 & $-9.74$ & $-6.89$ & $-9.82$ & $\mathbf{-4.76}$ & $-9.70$ & $-9.39$ \\
\midrule
\multirow{3}{*}{Music}  & 0.9 & $-1.40$ & $\mathbf{-0.73}$ & $-4.94$ & $-0.91$ & $-2.53$ & $-1.13$ \\
                        & 0.8 & $-1.76$ & $\mathbf{-0.47}$ & $-4.51$ & $-0.59$ & $-2.10$ & $-1.32$ \\
                        & 0.5 & $-2.64$ & $\mathbf{-1.31}$ & $-5.51$ & $-1.44$ & $-4.89$ & $-2.24$ \\
\bottomrule
\end{tabular}
\end{table}

The wide pair removes the initialisation confound, because both of its students start from the same random checkpoint. On this pair the speech cells are near-ties, for example $-1.38$ against $-1.56$ pp at $\lambda_1{=}0.8$, and the music cells clearly favour \Lcl{}, $-1.32$ against $-2.10$ pp at the same weight. The ordering therefore survives without the HuBERT initialisation. On speech, however, the margin is too small to attribute the robustness gain to the loss alone. On the merged scores the wide \Lcl{} pair leads by $10.8$ Average Score points at $\lambda_1{=}0.9$ ($931.9$ against $921.1$). At three layers and $\lambda_1{=}0.8$, \Lcl{} reaches $959.4$ Speech Score against $924.1$, with IC $93.57\%$ against $89.24\%$ and ASR $13.62\%$ against $17.08\%$ word error rate (WER).

\begin{table*}[t]
\centering
\caption{Per-task results for the two-layer students ($D{=}768$, HuBERT initialisation) under both merging algorithms. The rows $\lambda_1{=}1.0$ and $0.0$ are the unmerged students, from which the $\Delta$ of~(\ref{eq:delta}) is measured; the ensemble baseline is $\lambda$-invariant. Scores follow~(\ref{eq:score}) and bold marks the higher Speech Score in each pair.}
\label{tab:pertask}
\scriptsize
\setlength{\tabcolsep}{3.5pt}
\begin{tabular}{@{}llcccccccccccc@{}}
\toprule
& & \multicolumn{4}{c}{\textit{Speech}} & \multicolumn{5}{c}{\textit{Music \& audio}} & \multicolumn{3}{c}{\textit{Scores}} \\
\cmidrule(lr){3-6}\cmidrule(lr){7-11}\cmidrule(lr){12-14}
$\lambda_1$ & Loss & ASR & KS & IC & ER & SingerID & VocID & InstCls & GenreID & ESC-50 & Sp.Sc & Mu.Sc & Avg \\
 & & {\scriptsize WER$\downarrow$} & {\scriptsize Acc$\uparrow$} & {\scriptsize Acc$\uparrow$} & {\scriptsize Acc$\uparrow$} & {\scriptsize Acc$\uparrow$} & {\scriptsize Acc$\uparrow$} & {\scriptsize Acc$\uparrow$} & {\scriptsize Acc$\uparrow$} & {\scriptsize Acc$\uparrow$} & {\scriptsize $\uparrow$} & {\scriptsize $\uparrow$} & {\scriptsize $\uparrow$} \\
\midrule
\multicolumn{14}{@{}l}{\textit{Ensemble distillation (reference)}} \\
-- & \Lkd{} & $15.14$ & $94.19$ & $85.55$ & $60.76$ & $73.07$ & $70.65$ & $64.09$ & $63.22$ & $72.35$ & $901.1$ & $954.2$ & $927.7$ \\
\midrule
\multicolumn{14}{@{}l}{\textit{Task arithmetic (TA)}} \\
\multirow{2}{*}{$1.0$} & \Lkd{} & $13.18$ & $96.04$ & $94.70$ & $61.49$ & $69.04$ & $64.30$ & $63.62$ & $63.45$ & $70.65$ & $944.1$ & $920.8$ & $932.5$ \\
 & \Lcl{} & $12.98$ & $96.11$ & $94.52$ & $63.92$ & $69.61$ & $62.46$ & $61.43$ & $66.21$ & $70.05$ & $\mathbf{964.9}$ & $916.6$ & $940.7$ \\
\cmidrule(lr){1-14}
\multirow{2}{*}{$0.9$} & \Lkd{} & $13.97$ & $95.92$ & $93.56$ & $62.36$ & $71.18$ & $66.96$ & $62.90$ & $63.40$ & $70.33$ & $945.5$ & $930.4$ & $938.0$ \\
 & \Lcl{} & $13.86$ & $96.04$ & $93.97$ & $62.88$ & $69.09$ & $63.32$ & $62.79$ & $65.17$ & $70.90$ & $\mathbf{951.8}$ & $921.2$ & $936.5$ \\
\cmidrule(lr){1-14}
\multirow{2}{*}{$0.8$} & \Lkd{} & $20.25$ & $94.46$ & $90.72$ & $64.79$ & $71.81$ & $66.15$ & $64.31$ & $61.72$ & $69.00$ & $935.2$ & $925.8$ & $930.5$ \\
 & \Lcl{} & $15.81$ & $95.75$ & $92.05$ & $62.40$ & $68.71$ & $64.45$ & $63.06$ & $64.83$ & $71.50$ & $\mathbf{935.7}$ & $924.9$ & $930.3$ \\
\cmidrule(lr){1-14}
\multirow{2}{*}{$0.5$} & \Lkd{} & $28.67$ & $93.51$ & $74.03$ & $61.24$ & $68.64$ & $63.75$ & $65.14$ & $61.72$ & $69.31$ & $830.6$ & $914.4$ & $872.5$ \\
 & \Lcl{} & $21.04$ & $94.19$ & $78.72$ & $62.15$ & $67.66$ & $61.70$ & $64.87$ & $63.45$ & $70.70$ & $\mathbf{876.1}$ & $914.0$ & $895.1$ \\
\cmidrule(lr){1-14}
\multirow{2}{*}{$0.0$} & \Lkd{} & $23.21$ & $89.16$ & $59.58$ & $58.14$ & $75.00$ & $66.29$ & $61.43$ & $66.21$ & $67.90$ & $767.7$ & $935.1$ & $851.4$ \\
 & \Lcl{} & $22.18$ & $89.65$ & $58.08$ & $58.93$ & $71.93$ & $64.46$ & $62.26$ & $61.34$ & $67.92$ & $\mathbf{774.7}$ & $911.0$ & $842.9$ \\
\midrule
\multicolumn{14}{@{}l}{\textit{Correlation-permutation (CP)}} \\
\multirow{2}{*}{$0.9$} & \Lkd{} & $14.23$ & $95.91$ & $93.92$ & $63.16$ & $70.01$ & $65.76$ & $63.78$ & $61.58$ & $71.15$ & $952.5$ & $923.9$ & $938.2$ \\
 & \Lcl{} & $13.52$ & $95.91$ & $93.99$ & $63.43$ & $70.01$ & $64.79$ & $62.00$ & $64.83$ & $71.06$ & $\mathbf{957.1}$ & $924.6$ & $940.9$ \\
\cmidrule(lr){1-14}
\multirow{2}{*}{$0.8$} & \Lkd{} & $16.77$ & $95.33$ & $90.30$ & $61.96$ & $70.67$ & $64.20$ & $65.26$ & $61.58$ & $69.20$ & $923.0$ & $920.5$ & $921.7$ \\
 & \Lcl{} & $15.06$ & $95.52$ & $90.43$ & $62.45$ & $70.31$ & $64.73$ & $62.01$ & $64.48$ & $71.15$ & $\mathbf{933.1}$ & $924.6$ & $928.9$ \\
\bottomrule
\end{tabular}
\end{table*}

Table~\ref{tab:pertask} locates the difference. At $\lambda_1{=}0.8$ the two-layer \Lkd{} student loses $7.07$ pp of ASR, $3.98$ pp of IC and gains $3.30$ pp of ER against its own endpoints, a speech mean of $-2.33$ pp. \Lcl{} loses less on ASR, KS and IC ($-2.83$, $-0.36$ and $-2.47$ pp) and $1.52$ pp on ER, for a mean of $-1.80$ pp. In both cases the degradation is spread across tasks and is largest on ASR; no single task collapses. \Lcl{} holds the higher Speech Score in all twelve interpolated cells, $\lambda_1 \in \{0.9, 0.8, 0.5, 0.1\}$ at three sizes, reversing only at the wide pair's music-only endpoint, by $0.9$ points. The \Lcl{} students are also the stronger pair before merging, which is why Table~\ref{tab:ta_robust} is endpoint-relative.

Per-task differences were tested on paired per-sample predictions from single runs, with McNemar's test for classification, a paired bootstrap for ASR and Benjamini--Hochberg correction. The \Lcl{} advantage is significant on ASR at $\lambda_1{=}0.8$ and $0.5$ for the two-layer pair, and on IC at the same two weights for the three-layer pair; the composite Scores, the mean $\Delta$ and the permutation statistics carry no test.

\section{Analysis of \Lcl{} on Permutation Merging}
\label{sec:cp}

We now ask whether the same effect appears under a mechanistically unrelated merging algorithm. TA never permutes and assumes a shared initialisation, whereas CP permutes activations of independently trained encoders and ignores the initialisation. If the loss effect were a TA artefact, CP should be indifferent to it.

Students that are already similar need fewer channels moved by~(\ref{eq:cp}), and Table~\ref{tab:ffn} gives the fraction moved at each merge point. The matched comparison is the three-layer pair, where students, calibration data and algorithm are identical and only the loss differs. The first-block FFN rate falls from $98.2\%$ under \Lkd{} to $78.3\%$ under \Lcl{} and the second from $99.3\%$ to $89.8\%$, while the attention rates drop less. By the third block both need near-complete FFN reordering ($99.5\%$ against $98.5\%$), so the advantage sits where the student FFNs first diverge.

\begin{table}[t]
\centering
\caption{Percentage of channels CP permutes at each merge point; lower means the students already share more structure. The wide \Lcl{} pair also differs in width, head structure and initialisation; the different-initialisation column shows what unaligned students look like.}
\label{tab:ffn}
\scriptsize
\setlength{\tabcolsep}{2.5pt}
\begin{tabular}{@{}lccccc@{}}
\toprule
 & \multicolumn{3}{c}{2-layer} & \multicolumn{2}{c}{3-layer} \\
\cmidrule(lr){2-4}\cmidrule(lr){5-6}
Merge point & \Lkd{} & wide \Lcl{} & \Lkd{} & \Lkd{} & \Lcl{} \\
Initialisation & shared & shared & different & shared & shared \\
\midrule
CNN 1 & $41.2$ & $40.2$ & $34.4$ & $42.0$ & $\mathbf{21.3}$ \\
CNN 2--3 & $0.0$ & $0.0$ & $99.0$--$99.6$ & $0.0$ & $0.0$ \\
CNN 4--5 & $7.6$--$28.9$ & $0.4$--$2.1$ & $99.4$--$99.6$ & $10.9$--$33.6$ & $\mathbf{0.0}$ \\
CNN 6--7 & $70.1$--$85.7$ & $40.8$--$68.2$ & $99.6$--$100$ & $72.1$--$84.0$ & $\mathbf{37.3}$--$\mathbf{68.2}$ \\
\midrule
T1 attn / FFN & $99.6/98.5$ & $100/99.9$ & $100/100$ & $99.0/98.2$ & $\mathbf{94.4/78.3}$ \\
T2 attn / FFN & $100/99.5$ & $100/100$ & $100/100$ & $99.9/99.3$ & $\mathbf{99.0/89.8}$ \\
T3 attn / FFN & -- & -- & -- & $100/99.5$ & $100/98.5$ \\
\bottomrule
\end{tabular}
\end{table}

The convolutional rows of Table~\ref{tab:ffn} point the same way. In the first convolutional layer, the three-layer \Lcl{} students need $21.3\%$ of the channels reordered against $42.0\%$ under \Lkd{}, and in layers four and five they need none against up to $33.6\%$. The two \Lcl{} students therefore keep more of the shared front end they start from. This may also explain why they score slightly lower on the music tasks, since a front end that stays close to the other student also stays close to the speech-initialised starting point.

The quality of the alignment agrees with its extent. We measure quality as the mean matched $|\rho|$, the average absolute correlation between the pairs of channels that CP matches. It is $68.6\%$ for the two-layer \Lcl{} students against $60.3\%$ for \Lkd{}, and $69.3\%$ against $59.9\%$ at three layers. Neither the permutation rates nor the matched correlations involve a downstream probe, so both are properties of the encoders themselves. For reference, replacing the shared initialisation by independent ones drops the two-layer \Lkd{} mean from $60.3\%$ to $47.0\%$, a larger change than the loss produces in the other direction.

Under CP \Lcl{} holds the higher Speech Score at both weights and depths ($957.1$/$952.5$ and $933.1$/$923.0$ at two layers, $964.6$/$958.5$ and $953.9$/$927.8$ at three). The Average Score follows in three of the four, the exception being three layers at $\lambda_1{=}0.9$ ($939.2$/$941.4$). The $\Delta$ of~(\ref{eq:delta}) on the endpoint rows of Table~\ref{tab:pertask} is more mixed. \Lcl{} is more robust on music ($-0.44$, $-0.45$ against $-1.90$, $-2.17$ pp), level on speech at $0.8$ ($-2.06$ pp) and behind at $0.9$ ($-0.07$ against $-0.44$). Intent classification is level at two layers ($90.30\%$ against $90.43\%$ at $\lambda_1{=}0.8$) and separates at three ($87.64\%$ against $93.20\%$). It stays within $87.6$--$94.9\%$ across all shared-initialisation pairs and both algorithms, the largest drop between weights being $5.24$ pp (three layers, \Lkd{}, CP). No configuration therefore degrades sharply on one task. Section~\ref{sec:converge} turns to the two algorithms together.

\section{Consistency Across Merging Algorithms}
\label{sec:converge}

Sections~\ref{sec:ta} and~\ref{sec:cp} are not conclusive alone, because \Lcl{} might suit the way one algorithm combines models. TA works in weight space and CP in activation space, so a benefit under both is harder to dismiss as a quirk of either. This section asks how far they agree.

\begin{figure}[t]
\centering
\includegraphics[width=0.92\columnwidth]{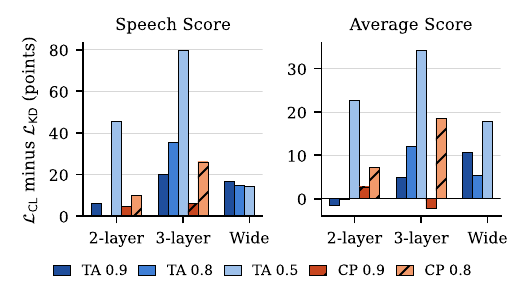}
\caption{Difference between the two losses, \Lcl{} minus \Lkd{}, in Speech Score (left) and Average Score (right); positive bars favour \Lcl{}. TA is shown at $\lambda_1 \in \{0.9, 0.8, 0.5\}$ and CP at $\{0.9, 0.8\}$, the weights at which CP was run. The wide configuration has no CP bars, since its \Lkd{} pair was not merged with CP.}
\label{fig:gap}
\end{figure}

Fig.~\ref{fig:gap} plots the difference between the two losses under both algorithms. \Lcl{} holds the higher Speech Score in all thirteen comparisons, nine under TA and four under CP, and the higher Average Score in ten. The exceptions are the two-layer TA merges at $\lambda_1{=}0.9$ and $0.8$ ($1.5$ and $0.2$ points to \Lkd{}) and the three-layer CP merge at $0.9$ ($2.2$ points). The largest gap is $79.5$ Speech Score points, at three layers and $\lambda_1{=}0.5$.

\begin{figure*}[t]
\centering
\includegraphics[width=0.82\textwidth]{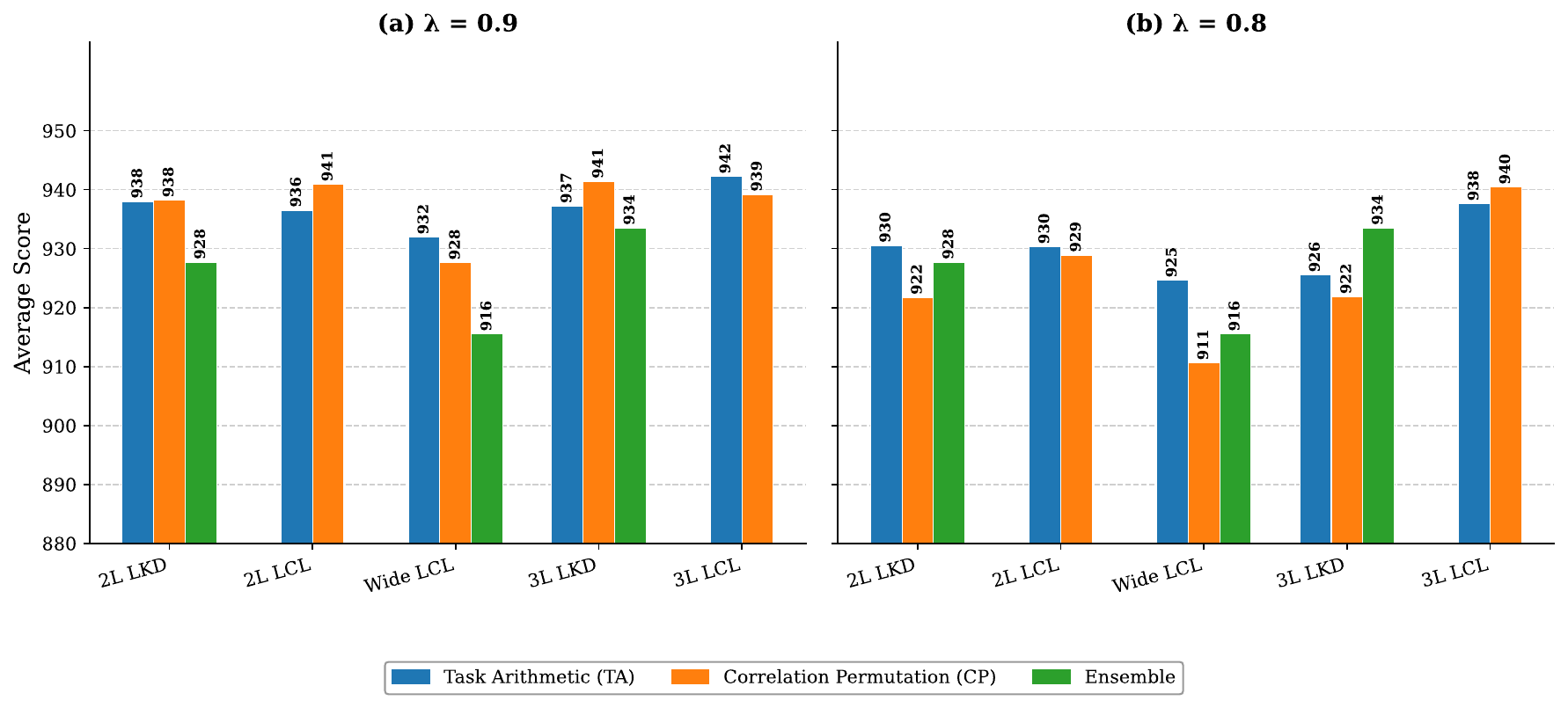}
\caption{Average Score~(\ref{eq:score}) for correlation-permutation (CP), task arithmetic (TA) and the ensemble-distillation baseline across five configurations at $\lambda_1{=}0.9$ and $\lambda_1{=}0.8$. The two post-hoc methods track each other closely at $\lambda_1{=}0.9$. At $\lambda_1{=}0.8$ two configurations separate, the wide pair and the two-layer \Lkd{} pair, where TA exceeds CP by $14.0$ and $8.8$ points (Section~\ref{sec:practice}).}
\label{fig:converge}
\end{figure*}

Fig.~\ref{fig:converge} compares the two algorithms. At $\lambda_1{=}0.9$ the CP-minus-TA Average Score ranges from $-4.3$ to $+4.4$ points, mean $+0.3$, with CP ahead on three of the five. At $0.8$ TA leads on four of the five by $5.0$ points on average, the wide pair by $14.0$ and the two-layer \Lkd{} pair by $8.8$. Of the $74$ testable per-task comparisons $12$ reach significance, all with $|h| < 0.13$ on classification tasks, and four of five configurations show none at $0.9$. The other $16$ of $90$ lacked prediction files.

This agreement points to the representation. Because TA and CP share no machinery, the only factor that can carry an effect across both is the student representation, which is what the loss shapes~\cite{what-transfer-learning}. They do share the probe suite used to score them, so part of the agreement could come from the evaluation. The rates and matched correlations of Section~\ref{sec:cp}, however, involve no probe, so the structural part of the effect belongs to the encoders. We hypothesise that \Lcl{}'s self-correlation term, Eq.~(\ref{eq:lsc}), drives it by decorrelating feature dimensions, which would flatten TA's path and ease CP's alignment. That hypothesis has two parts, and only the first is established. A companion ablation of the same objective~\cite{ritter_mechanistic_2026} shows that the self-correlation term does reorganise the feature space, redistributing information across dimensions rather than removing it, and that this alone improves clean-condition intent classification and emotion recognition. Whether that reorganisation is what makes the students merge better remains untested, since that work performs no merging. A direct test would sweep $\lambda_{\text{sc}}$ and re-merge at each value, which we leave to future work.

\section{Discussion}
\label{sec:practice}

A practitioner must still choose a merging method, and Table~\ref{tab:tradeoffs} compares the three by their practical requirements. Task arithmetic is cheapest at merge time but requires a shared initialisation; CP removes that constraint at the cost of calibration data and minutes of compute; ensemble distillation fixes $\lambda$ at training time. \Lcl{} improves both post-hoc methods on the speech side, so the choice turns on the practical constraints. The gain is on the speech side. At standard width the \Lkd{} students keep the higher Music Score at every weight, consistent with their stronger music endpoint ($935.1$ against $911.0$ at $\lambda_1{=}0.0$), so \Lkd{} remains the better choice when music performance is the priority.

\begin{table}[t]
\centering
\caption{Operating points of the three merging methods.}
\label{tab:tradeoffs}
\footnotesize
\setlength{\tabcolsep}{4pt}
\begin{tabular}{@{}llll@{}}
\toprule
Aspect & Ensemble distill. & Task arithmetic & Corr.-perm. \\
\midrule
Training        & Joint distill.   & Two runs        & None \\
Shared init     & N/A              & Required        & Not required \\
Post-hoc $\lambda$ & No            & Yes             & Yes \\
Model size      & Any              & Compressed      & Any \\
Calibration     & Training data    & Not needed      & $\sim$10K samples \\
Merge compute   & GPU-days         & Seconds         & Minutes \\
\bottomrule
\end{tabular}
\end{table}

At $\lambda_1{=}0.9$ all ten post-hoc merges beat the ensemble baseline of their own architecture (Fig.~\ref{fig:converge}). At $0.8$ four fall below it, the two-layer \Lkd{} pair under CP, the three-layer \Lkd{} pair under both algorithms and the wide \Lcl{} pair under CP. Both standard-width \Lcl{} pairs stay above. Interpolation also degrades more slowly at $0.8$, the two-layer \Lkd{} pair falling $938.0$ to $930.5$ under TA against $938.2$ to $921.7$ under CP, roughly twice as far.

The clearest disagreement between the two algorithms is the wide randomly-initialised configuration ($D{=}1536$), where TA exceeds CP by $4.3$ points at $\lambda_1{=}0.9$ and by $14.0$ points at $0.8$. Two factors plausibly favour TA here. First, Entezari et al.~\cite{permutation_linear_interpolation} conjecture that the loss barrier under linear interpolation decreases as hidden width grows; the argument is made for two-layer ReLU networks and is not established for transformers. Second, the wide student has $6$ attention heads with $d_k{=}256$ instead of $12$ heads with $d_k{=}64$. CP first reorders whole heads and then permutes the channels within each head, so on the wide student the head-ordering step chooses among $6$ heads instead of $12$, and each within-head assignment spans $256$ channels instead of $64$. The alignment CP can find is therefore coarser. This is not because the wide students were already aligned; Table~\ref{tab:ffn} shows that this pair needs $99.9$--$100\%$ transformer permutation, as much as any configuration in the table.

CP's binding constraint is the initialisation, not the loss. Two independently initialised \Lkd{} students, which task arithmetic cannot merge, reach $933.8$ Average Score at $\lambda_1{=}0.9$ against $938.2$ for the shared pair. At $0.8$ they fall to $815.5$, IC dropping from $93.99\%$ to $65.49\%$, so permutation buys freedom from shared initialisation only in the speech-biased regime. Whether the loss effect transfers to TIES-Merging~\cite{tiesmerging}, DARE~\cite{dare} or speech-specific merging~\cite{speechft} remains open.

\section{Conclusion}
\label{sec:conclusion}

The distillation loss is not merge-neutral. Replacing \Lkd{} with \Lcl{} at fixed architecture and evaluation leaves distilled speech-music students easier to merge under both algorithms. Under interpolation they stay closer to their own better endpoint in $17$ of $18$ comparisons and hold the higher Speech Score at every interpolated weight; under permutation the three-layer pair needs fewer channels rearranged, both depths match more strongly, and speech scores are again higher. \Lkd{} keeps the higher Music Score at standard width. Because the two algorithms share no mechanism yet agree, the effect is more plausibly a property of the representation than of either algorithm, so the loss should be chosen with merge-time behaviour in mind.

\balance
\printbibliography

\end{document}